\documentstyle[preprint,aps]{revtex}

\begin{document}

\preprint{ January 1995}

\title{\bf Kazakov--Migdal Model with Logarithmic Potential and the
Double Penner Matrix Model}

\author{Lori Paniak and Nathan Weiss}
\address{Department of Physics, University of British Columbia,\\ Vancouver,
British Columbia, Canada V6T 1Z1}

\maketitle
\vglue -.2in
\begin{abstract}
The Kazakov--Migdal (KM) Model is a U(N) Lattice Gauge Theory with  a
Scalar Field in the adjoint representation
but with no kinetic term for the Gauge Field.
This model is formally soluble in the limit $N\rightarrow \infty$
though explicit solutions
are available for a very limited number of scalar potentials.
A ``Double Penner''
Model in which the potential has two logarithmic singularities
provides an example of a  explicitly soluble model. We begin
by reviewing the formal
solution to this Double Penner KM Model. We pay special
attention to the relationship of this model to
an ordinary (one) matrix model whose potential has two logarithmic
singularities (the Double Penner Model).
We  present a
detailed analysis of the large N behavior
of this Double Penner Model. We describe the
various one cut and two cut solutions and we discuss cases in
which ``eigenvalue condensation'' occurs at the singular points of
the potential.
We then describe the consequences of our study for the
KM Model described above. We present the phase diagram
of the model and describe its critical regions.

\end{abstract}
\pacs{ }
\newpage
\section{Introduction}
Several years ago Kazakov and Migdal \cite{kmm} proposed a model which
they hoped would provide a description of Quantum Chromodynamics
(QCD) in the limit in which the
number $N$ of colors is large. Their model is a Lattice Field Theory
defined on a $D$ dimensional
hypercubic lattice with sites labeled as $x$ and links labeled by
pairs of nearest neighbor sites $<x,y>$. The fields in the model
are a scalar field $\phi(x)$ which, for every site $x$, is an
$N\times N$ Hermetian matrix and a Gauge field $U(x,y)$ which,
for every link $<x,y>$ of the lattice, is an $N\times N$ Unitary
matrix. The Action for a given configuration $\phi(x)$ and
$U(x,y)$ in their model is given by
\equation
	S_{\rm KM}=N\sum_x{\rm
	Tr}V[\phi(x)]-N\sum_{<x,y>}{\rm
	Tr}\left(\phi(x)U(x,y)\phi(y)U^{\dagger}(x,y)\right)
	\label{skm}
\endequation
where $V(\phi)$ is a potential function for the scalars which is,
at this stage, arbitrary. The factors of $N$ are included to assure
a smooth large $N$ limit for the model.
The model is defined by the Partition Function
$Z_{\rm KM}$ given by the following Functional Integral:
\equation
	Z_{\rm KM}=\int \prod_x d\phi(x) \prod_{<x,y>}
	[dU(x,y)] \exp\left[-S_{\rm KM}\right]
	\label{km}
\endequation
Here $d\phi(x)$ is the Hermetian integration measure
over the matrix $\phi(x)$ and $[dU(x,y)]$ is the invariant Haar
measure for integration of the matrix $U(x,y)$ over the unitary
group $U(N)$.

This model is invariant
under the gauge transformations
\equation
	\phi(x)\rightarrow
	\omega(x)\phi(x)\omega^{\dagger}(x);~~~~ U(xy)\rightarrow
	\omega(x)U(xy)\omega^{\dagger}(y)
	\label{gaugetrans}
\endequation
where $\omega(x)$ is an arbitrary
$U(N)$ valued function of $x$.
The second term in Eq. (\ref{skm}) is the usual
gauge invariant kinetic term for
a scalar field in the adjoint representation of a gauge group.
In fact this model (which is called the Kazakov--Midgdal Model or
the KM Model) is simply a model of an adjoint scalar coupled to
a Gauge Field $except$ that the usual kinetic term for the Gauge Field
is omited.
It is the absence of this term that makes the model soluble in
the large $N$ limit. We begin by reviewing how this works.

The first step, which can be done for finite $N$, is to explicitly
integrate over all the unitary matrices $U(x,y)$ in Eq. (\ref{km}).
This can be done since there is no coupling between the various
$U(x,y)$. The result is a Functional Integral over the fields
$\phi(x)$ which. As a result of the Gauge Invariance (\ref{gaugetrans}) the
integral depends only on
the eigenvalues $\phi_i(x)$ ($i=1..N$) of the
matrices $\phi(x)$. In fact the
 integral can be done explicitly using the  ``Itzykson--Zuber''
formula \cite{izform}. (For details see Ref. \cite{kmm}.) The result is:
\equation
	Z_{\rm KM}\propto\int\prod_{x,i} d\phi_i(x)\Delta^2[\phi(x)]
	\exp \left[{-N\sum_x {\rm Tr}V[\phi(x)]}\right]\prod_{<x,y>}
	{\det_{ij}e^{N\phi_i(x)\phi_j(y)}\over\Delta[\phi(x)]\Delta[\phi(y)]}
	\label{phii}
\endequation
where $d\phi_i(x)$ is the ordinary integration measure over the real numbers
$\phi_i(x)$ and
$\Delta[\phi]=\det_{ij}(\phi_i)^{j-1}=\prod_{i<j}(\phi_i-\phi_j)$ is
the Vandermonde determinant for $\phi$.

In this form of the Functional Integral it is well known \cite{largen}
how to go to the limit of large $N$.
When $N\rightarrow \infty$ the partition function (\ref{phii})
is dominated by the stationary points of the action (defined as minus the
logarithm of the integrand).In such circumstances the stationary
points ($\Phi_i(x)$) are called the Master Fields for the Theory. The value
of the integral when $N\rightarrow\infty$ is equal to the integrand
evaluated when $\phi_i(x)$ is equal to the Master Field $\Phi_i(x)$.

When $N$ is infinite it is customary and convenient to describe the
Master Field $\Phi_i$ will will usually be independent of $x$ by
a density of eigenvalues. The idea is to order the eigenvalues so that
$\Phi_i$ is monotonically increasing and then to define the density of
eigenvalues
\equation
	\rho(\lambda)=\frac{1}{N}
	\left(\frac{d\Phi}{di}\right)^{-1}({{\Phi_i=\lambda}})
	\label{rhodef}
\endequation
so that
\equation
	\int_{-\infty}^{\infty} \rho(\lambda)d\lambda =1
	\label{rhonorm}
\endequation
$N\rho(\lambda)d\lambda$ is equal to the number of eigenvalues in a
range $d\lambda$ about $\lambda$.

The simplest KM Model in which the potential
$V(\phi)=m^2\phi^2/2$ is quadratic was first solved
by Gross \cite{gross}. There is a wealth of literature both
on the solutions to the saddle point equations
and on the relationship of the KM Model with QCD. A selection of references
are Refs. \cite{mig} - \cite{kazak}.

It was originally thought that this model would have a second order phase
transition
and some evidence was given in the case of
a quadratic potential that this  occurs when $m^2=2D$.
It was then argued that the critical behavior of this model should be
represented by QCD, the only known nontrivial four dimensional field
theory with non-Abelian gauge symmetry. Unfortunately the solution of
Gross \cite{gross} showed that the Gaussian model had no critical behavior.
This problem, combined with a
further problem of an additional local $Z_N$ symmetry which implies the
vanishing of Wilson Loops \cite{ksw} and a better understanding
of why the Gaussian Model fails to induce QCD has led to a consensus that
the KM Model does not induce QCD.

Nonetheless the model (\ref{km}) is interesting in its own right both
as a Gauge Theory which is soluble in large $N$ and, as we shall review
below, as an interesting example of a Matrix Model.
Fortunately there has recently \cite{dms},\cite{makpen} been some progress
in finding an explicit solution to a non-Gaussian Kazakov-Migdal
Model with a logarithmic potential of a very specific form.
This solution was found by relating the
density of eigenvalues for the KM  Model to that of the
an ordinary  (one) Matrix Model whose potential has two logarithmic
 singularities.  Although much is known about
Matrix Models for polynomial potentials and for  potentials with
a $single$ logarithmic singularity  (the Penner Model)
little is know about the model with two logarithmic
singularities (which might be called a ``Double Penner'' Model).

The purpose of this paper is twofold. First of all to study  ordinary
Matrix Models with a ``Double Penner'' Potential (which has two
logarithmic singularities) and then to relate the solution
of this problem to the solution of the Kazakov Migdal Model with
a specific class of logarithmic potentials.

The results of this paper are applicable in a much wider context
than just the KM Model. If, for example, we choose the dimension
$D=1/2$ we recover the solution of the ordinary ``Two Matrix Model''
\cite{makpen}. The techniques discussed in this paper and in Ref. \cite{dms}
are  applicable to ordinary Matrix Models whose wide range of
applicability can be seen in Refs. \cite{mehta} and \cite{twodgrav}.

The plan of the paper is as follows. In Section \ref{method}{} we review the
method of \cite{dms} of finding the eigenvalue distributions for
both Kazakov--Migdal Models and ordinary Matrix Models. We then
show in general how this allows for explicit solutions in the
case of a `Double Penner'' potential and how the
density of eigenvalues for the KM Model is related to that
of an ordinary matrix model. In Section \ref{s3} we describe in detail
the variety of solutions to the ordinary Matrix
Model with a Double Penner potential. We discuss the various
one cut and two cut solutions and the cases in which there is
``condensation of eigenvalues'' at the singularities.
In Section \ref{s4} we apply the results of Section \ref{s3} to the
Double Penner KM Model. We present a phase diagram for the model
and review its critical behavior.  In Section \ref{s5} we
summarize our results and conclusions.

\section{Method of Solution} {\label{method}{}}

\subsection{Ordinary Matrix Model}

Consider first an ordinary Hermitian Matrix Model whose only variable is a
single $N\times N$ Hermetian Matrix $\phi$. The model is
defined by  by the partition function
\equation
	Z=\int d\phi~~ {\rm exp}\left[-N{\rm tr}V(\phi)\right]
	\label{met1}
\endequation
where $d\phi$ is the Hermetian integration measure. If (as is always
assumed) Tr$V(\phi)$ is invariant under $U(N)$ transformations of $\phi$
then the integrand in Eq. (\ref{met1}) depends only on the eigenvalues
$\phi_i$ of $\phi$ and the partition function can be written as
\equation
	Z\propto\int \prod_i d\phi_i~~
	\Delta^2(\phi)~~ {\rm exp}\left[-N{\rm tr}V(\phi)\right]
	\label{met2}
\endequation
As discussed in the Introduction,
in the limit $N\rightarrow \infty$ this integral is dominated
by a matrix $\Phi$ whose eigenvalues $\mu$ are distributed
according to some distribution $\rho(\mu)$.

There are many methods of studying and solving such Matrix Models.
The method we discuss here is most suitable for generalization to
the KM Model \cite{M92}.
The density of eigenvalues can be found by defining the quantity
\equation
	E_\lambda=\langle{1\over N}{\rm Tr}{1\over {\lambda -\phi}}\rangle
	\equiv {1\over Z}\int d\phi~~\left({1\over N}
	{\rm Tr}{1\over {\lambda -\phi}}
	\right)
	{\rm exp}\left[-N{\rm tr}V(\phi)\right]
\endequation
where $\lambda$ is an arbitrary complex number. Once $E_\lambda$ has been
computed
the density of eigenvalues can  easily be determined since at large $N$
\equation
	E_\lambda~=~{1\over N}{\rm Tr}{1\over{\lambda-\Phi}}~=~
	\int_{-\infty}^{\infty}d\eta ~\rho(\eta){1\over{\lambda-\eta}}
\endequation
So that
\equation
	2\pi i~\rho(\eta)~=~ E_{\lambda-i\epsilon}-E_{\lambda +i\epsilon}
\endequation
which is nonzero only along the branch cuts of $E_\lambda$. Thus by finding
the branch cuts of $E_\lambda$ and computing the discontinuities across
these cuts we can compute the density of eigenvalues $\rho(\lambda)$.

Following Ref. \cite{dms} we begin with the equation
\equation
	\int d\phi~~{d\over {d\phi_{ij}}}\left\{\left( {1\over{\lambda
	-\phi}}\right)_{ij} {\rm exp}\left[-N{\rm tr}V(\phi)\right]
	\right\}~=~0
	\label{loop1}
\endequation
This is now written in terms of an integral over the $matrices$ $\phi$ rather
than in tems of their eigenvalues.
Eq. (\ref{loop1}) leads to the following equation for $E_\lambda$
\equation
	E_\lambda^2-\langle{1\over N}{\rm Tr}
	{1\over{\lambda-\phi}}V^\prime(\phi)\rangle
	~=~0
	\label{onemme}
\endequation
In the simplest case in which $V(\phi)=m^2\phi^2/2$ this equation
leads immediately to a quadratic equation for $E_\lambda$
\equation
	E_\lambda=\frac{1}{2}\left(m^2\lambda-
	\sqrt{m^4\lambda^2-4m^2}\right)
\endequation
and thus to the well known semicircle distribution of the eigenvalues of $\phi$
\equation
	\rho(\lambda)=\frac{m}{\pi}\sqrt{1-\frac{m^2\lambda^2}{4}}
\endequation
which has support on the interval $(-2/m,2/m)$.

\subsection{Kazakov--Migdal Model \label{sskmm}}

The situation for the Kazakov--Migdal Model is significantly more
complicated mostly due to the presence of the gauge field.
One option for solving this model is to begin with
the Functional Integral (\ref{km}) and to define the two quantities
$E_\lambda$ and $G_{\lambda \mu}$ as follows: Let $a$ and $b$ be
two adjacent sites on the lattice and let $U_{ab}$ be the gauge field
on the link joining them then define the average $<Q>$ of any quantity $Q$
as
\equation
	<Q>=\frac{1}{Z_{KM}}\int \prod_x d\phi(x) \prod_{<x,y>}
	[dU(x,y)]~Q~ \exp\left[-S_{\rm KM}\right]
\endequation
We then define
\equation
	E_\lambda=
	\langle{1\over N}{\rm Tr}{1\over{\lambda-\phi(a)}}\rangle
\endequation
which is expected to be independent of the chosen site $a$
(in the limit of large volume) and
\equation
	G_{\lambda\mu}=\langle{1\over N}{\rm tr} \left(
	{1\over{\lambda-\phi(a)}}U_{ab}
	{1\over{\mu-\phi(b)}}U_{ab}^{-1}\right)\rangle
	\label{gdef}
\endequation
which is expected to be independent of the chosen link $(a,b)$.
It is useful to note that asymptotically, for large $\lambda$,
\equation
	E_{\lambda} \sim  \frac{1}{\lambda} + \sum_{n=1}
	\frac{\langle 	\phi^{n} \rangle}{\lambda^{n+1}}
	\label{easym}
\endequation
\equation
	G_{\lambda \mu}  \sim  \frac{E_{\mu}}
	{\lambda} +... \label{gasym}
\endequation

The next step is to write two equations analogous to Eq. \ref{loop1};
one for $E_\lambda$ and one for $G_{\lambda\mu}$
\equation
	\int \prod_x d\phi(x) \prod_{<x,y>}
	[dU(x,y)]
	{d\over{d\phi(a)_{ij}}}\left\{\left( {1\over{\lambda
	-\phi(a)}}\right)_{ij}{\rm exp}~\left[-S_{KM}(\phi,U)\right]
	\right\}~=~0
	\label{elamkmm}
\endequation
and
\equation
	\int{\cal D}\phi{\cal D}U
	{d\over{d\phi(a)_{ij}}}\left\{\left( {1\over{\lambda
	-\phi(a)}}U_{ab}{1\over{\mu
	-\phi(b)}}U_{ab}^{-1}\right)_{ij}{\rm exp}
	~\left[-S_{KM}(\phi,U)\right]\right\}~=~0
	\label{gmunukmm}
\endequation
where	$S(\phi,U)$
is the Kazakov--Migdal Action (\ref{skm}).

Recall that in the limit of infinite
$N$ the functional integral is dominated by a single, translationally
invariant Master Field $\Phi$. Due to the Gauge Invariance of the
Action we can choose $\Phi$ to be diagonal with eigenvalues
$\Phi_i$ ($i=1..N$) without any loss of generality.

Before proceeding it is useful to define a quantity $\Lambda(\phi)$
as follows:
Consider first the quantity
\equation
	C_{ij}(\Phi)={1\over Z_{KM}}\int [dU] \vert U_{ij}\vert^2
	{\rm exp}\left(-S_{KM}(U,\Phi)\right)
\endequation
Then define
\equation
	\Lambda_i(\Phi)=C_{ij}(\Phi)\Phi_j
\endequation

With this definition in hand one can derive
the following equations
(see Ref. \cite{dms} for details) corresponding to Eqs. \ref{elamkmm}
and \ref{gmunukmm} respectively:
\equation
	E_\lambda^2-\langle{1\over N}{\rm
 	Tr}{{V^\prime(\Phi)-2D\Lambda(\Phi)}\over
	{\lambda-\Phi}}\rangle=0
	\label{eqq1}
\endequation
\equation
	(E_\lambda+\mu)G_{\lambda \mu}-E_\lambda-\langle{1\over N}{\rm Tr}
	{{V^\prime(\Phi)-(2D-1)\Lambda(\Phi)}\over
	{\lambda-\Phi}}U_{xy}{1\over{\mu-\Phi}}U_{xy}^{-1}
	\rangle = 0
	\label{eqq2}
\endequation
Note $\Lambda(\phi)$ is defined so that $\langle \cdot\cdot\cdot
U_{xy}\phi(y)U_{xy}^{-1}\rangle$ is replaced by
$\langle \cdot\cdot\cdot \Lambda(\Phi)\rangle$ in the large $N$
limit provided the full expression is gauge invariant and
that $U_{xy}$ does not appear in the expression again.
At this stage we must make an important comment about notation.
In Equation (\ref{eqq2}) the only average remaining is that over
the Gauge fields which is done in the background field $\Phi$.
After doing this average all that remains is to perform the trace
which can be thought of as an average over the distribution of
eigenvalues of $\Phi$. It is thus common to write, for example,
$V^\prime(\phi)$ or $\Lambda(\phi)$ instead of $V^\prime_i(\Phi)$
or $\Lambda_i(\Phi)$ where we identify the label $i$ with the
eigenvalue $\phi$ so that $\Phi_i=\phi$. In the case when
the potential is the trace of a function of the matrix scalar
field, $V^\prime(\phi)$ will have the same form as the derivative
of the potential with respect to the matrix valued field.

To proceed further we need to make a choice for the potential
$V(\phi)$. The solution to the above equations is simplest in the
case of a Gaussian potential which is discussed in detail in
Ref. \cite{dms}. In this case it turns out that $\Lambda(\phi)$
is simply proportional to $\phi$. As a result one obtains a quadratic
equation for $E_\lambda$ whose solution (see \cite{dms} and \cite{gross})
yields a semicircle  distribution of eigenvalues where the edge
of the distribution $2/M(m)$ is a calculable function of the parameter
$m$ in the potential.

 In this paper we shall discuss
the more complicated case of a ``Penner'' potential.
Previous work on this subject can be found in Ref. \cite{makpen}. The case
which can be solved involves choosing a potential for which
\equation
	V^\prime (\phi)-(2D-1)\Lambda(\phi)=
	 {q\over{\phi-\xi}}+B
\label{ansatz}
\endequation
which has a pole at $\phi=\xi$ with residue
$q$ and an asymptotic value of $B$ at $\phi\rightarrow \infty$.
The idea is to use this form of $V^\prime -(2D-1)\Lambda(\phi)$
and to solve for the potential $V$ which leads to this function.

Using Eqs. ({\ref{eqq1}) and (\ref{eqq2}) (see \cite{dms} for more
details) we can derive the following equation which relates
$G_{\lambda\mu}$ to $E_\lambda$ and $G_{\xi\mu}$ (which thus involves
$G$ at the singularity of $V$):
\equation
	G_{\lambda \mu} =\frac{(\lambda-\xi)E_\lambda-qG_{\xi\mu}}
	{(\lambda-\xi)(E_\lambda+\mu-B)-q}
	\label{eqforg}
\endequation
This can be written in a more useful form by using the asymptotic
condition for $G_{\lambda\mu}$.
If we expand $G_{\lambda\nu}$ in Eq. (\ref{eqforg}) in large $\lambda$
using Eq. (\ref{gasym})
we find  that
\begin{equation}
	qG_{\xi\mu} = 1+(B-\mu)E_\mu
\end{equation}
Substituting this back into Eq. (\ref{eqforg}) we relate $G_{\lambda\nu}$
to $E_\lambda$ and $E_\nu$:
\begin{equation}
	G_{\lambda \mu}=\frac{(\lambda-\xi)E_\lambda+(\mu-B)E_\mu-1}
	{(\lambda-\xi)(E_\lambda+\mu-B)-q}
	\label{geqat}
\end{equation}

The next step is to notice from Eq. (\ref{gdef}) that $G_{\lambda\nu}$ is
symmetric
under interchange of $\lambda$ and $\nu$ so that
$G_{\lambda\nu}=G_{\nu\lambda}$.
Applying this condition to Eq. (\ref{geqat}) one finds, after some algebra,
the following quadratic equation for $E_\lambda$:
$$
	(\lambda-B)\lambda-\xi)~{E}_\lambda^2+\left[(\lambda-B)(\lambda-\xi)(\xi-B)
	+q(B-\xi)-(\lambda-\xi)\right]~E_\lambda
$$
\equation
	 +~\lambda(B-\xi)~=~{\rm constant}
\endequation
where the constant is independent of $\lambda$.
It is possible to determine this constant in terms of the mean value
of $\phi$; $\bar\phi=\int \rho(\phi)\phi d\phi$ using the asymptotic condition
(\ref{easym}). The equation for $E_\lambda$ then becomes:
$$
	(\lambda-B)(\lambda-\xi)~E_\lambda^2+\left[(\lambda-B)(\lambda-\xi)(\xi-B)
	+q(B-\xi)-(\lambda-\xi)\right]~E_\lambda
$$
\equation
	 +\left[ (\lambda+\bar\phi)(B-\xi)+(\xi^2-B^2)\right]~=~0
	\label{quad2}
\endequation
In principle it is
necessary to determine the value of $\bar\phi$ self consistently
by extracting the density of eigenvalues $\rho$, which will depend
on $\bar\phi$, from Eq. (\ref{quad2}) and then demanding that
$\int d\phi \rho(\phi)\phi~=~\bar\phi$. We shall see however that in
many cases
a properly normalized solution will exist for range of values of $\bar\phi$.
The physical reason for this is that due to the singularities of the
potential the eigenvalues (in large $N$) can be arbitrarily distributed
among two minima on different sides of a singularity. This point will be
discussed
further later in this paper.

It is now possible to extract the functions $\Lambda(\phi)$ and the potential
$V(\phi)$ by comparing the equation (\ref{quad2}) with Equation (\ref{eqq1}).
Notice that if $V^\prime-2D\L ambda$ would have only simple poles we would
also obtain (from Eq. (\ref{eqq1}) a quadratic equation for $E_\lambda$.
By comparing these two equations one finds (see \cite{dms} for details)
\equation
	V^\prime(\lambda)-2D\Lambda(\lambda)~=~
	\frac{1-q}{\lambda-B}+\frac{q}{\lambda-\xi}+(B-\xi)
	\label{vprime2dl}
\endequation
Using the ansatz (\ref{ansatz}) we find
\equation
	\Lambda(\lambda)~=~\frac{q-1}{\lambda-B}+\xi
	\label{lambda}
\endequation
which leads finally to the expression for $V^\prime$
\equation
	V^\prime(\lambda)~=~\frac{q}{\lambda-\xi}
	+\frac{(2D-1)(q-1)}{\lambda-B}
	+(2D-1)\xi+B
	\label{vprime}
\endequation
which results from our ansatz (\ref{ansatz}). This now yields the
potential $V$ for which we have found a solution.

The preceding results thus establish an interesting relationship between the
KM Model with two logarithmic singularities as in Eq. (\ref{vprime}) and
a ordinary one matrix model with the potential $W(\phi)$ whose derivative
is of the general form
\equation
	W^\prime(\phi)=\frac{r_1}{\phi-\eta_1}+\frac{r_2}{\phi-\eta_2}+C
	\label{womm}
\endequation
with
\equation
	r_1+r_2~=~1;~~~~~\eta_2-\eta_1~=~C
	\label{wommc}
\endequation
where $r_1$ is identified with $q$, $\eta_1$ with $\xi$ and $\eta_2$ with $B$.
A Matrix Model with the  potential $W(\phi)$
will have the same function $E(\lambda)$ and thus, in particular, the
same distribution of eigenvalues as the KM Model. This One Matrix Model can
be discussed for arbitrary values of $r_1,r_2$ and $C$ (i.e. without
requiring the conditions (\ref{wommc}) ) and it
has many interesting aspects. For example it generalizes the Penner Model
which contains a single logarithmic singularity to a model with
two singularities and, as will be seen later, it admits a multi--phase
solution space with non--trivial critical behavior.  In Section \ref{s3}
we thus study the double penner One Matrix Model whose results will later
be applied to the KM Model.

\subsection{Interpretation of Solutions \label{interpret}}

Before proceeding it is useful to recall that the infinite $N$ behavior
of both the ordinary Matrix Model and the K--M Model can be described by
the solution to an analogue mechanical problem. In the case of the
{\bf ordinary} Matrix Model the partition function Eq. (\ref{met1}) is written
in terms
of the eigenvalues $\phi_i$ of the matrix $\phi$
$$
	Z=\int \prod_{i=1}^N d\phi_i~~\Delta^2(\phi)~
	{\rm exp}\left[-N\sum_i V(\phi_i)\right]
$$
\equation
	=\int \prod_{i=1}^N d\phi_i~~{\rm exp}\left[-N\sum_i V(\phi_i)
	+\sum_{i<j}{\log {(\phi_i-\phi_j)^2}}\right]
	\label{int1}
\endequation
In the large $N$ limit in which the solution is given by the classical extremum
of the action we see that we have an analogue mechanical problem of $N$
particles which are constrained to lie on a line at locations
$\phi_1\cdot\cdot\cdot \phi_N$.
Each particle is subjected to an overall
 potential $V(\phi_i)$ and to a logarithmically
repulsive two--body potential $\log{(\phi_i-\phi_j)^2}$.

The solution to the {\bf KM Model} corresponds to another, more complicated,
analogue mechanical model. If we look at the
partition function in Eq. (\ref{phii}) and recall that in the infinite $N$
limit the $\phi(x)$ are independent of $x$ and that on a square lattice
each site has $D$ independent
nearest neighbors (where $D$ is the dimensionality of
the spacetime) then for infinite $N$ Eq.(\ref{phii}) is equivalent to
solving the following one matrix model
$$
	Z=\int \prod_{i=1}^N d\phi_i~~\Delta^2(\phi)~
	{\rm exp}\left[-N\sum_i V(\phi_i)\right]
	\left( \frac{{\rm det}_{ij}{\rm e}^{N\phi_i\phi_j}}
	{\Delta^2{(\phi)}}\right)^{D}~~=~~~
	\int \prod_{i=1}^N d\phi_i
$$
\equation
	{\rm exp}\left[-N\sum_i \left(V(\phi_i)-D\phi^2\right)
	-(D-1)\sum_{i<j}{\log {(\phi_i-\phi_j)^2}}
	+D\log{
	{\rm det}_{ij}{\rm e}^{N\left(\phi_i-\phi_j\right)^2/2}}\right]
	\label{interp1}
\endequation
This is once more a problem of $N$ particles on a line at locations $\phi_i$.
Their central potential is now $V(\phi)-D\phi^2$ and their interaction
is no longer a two--body interaction since it involves a determinant.
We can, however, use the fact that the integral over the Gauge Group
in Eq. (\ref{km}) which lead to Eq. (\ref{phii}) is nonsingular when
any two eigenvalues approach each other. Using this we see that the
effective interaction is logarithmically repulsive at short distances
(due to the extra factor of $\Delta^2$) and attractive at long
distances if $D>1$.

The analogue mechanical problem presented in this section is a very
useful tool for visualizing and checking the solution we obtain
using the mathematical machinery of Matrix Models.

\section{The Double Penner Model} {\label{s3}}

\subsection{Formal Solution \label{ss3a}}
In this section we investigate the large $N$ solutions
of the (non--KM)
 One Matrix Model with the potential (\ref{womm}), for all possible values of
the parameters $r_1, r_2, \eta_1, \eta_2$ and $C$.
The basic equation is Eq. (\ref{onemme}) and leads to the quadratic equation
\equation
	E_\lambda^2-W^\prime(\lambda)E_\lambda
	+\frac{r_1E_{\eta_1}}{\lambda-\eta_1}
	+\frac{r_2E_{\eta_2}}{\lambda-\eta_2}~=~0
	\label{dp1}
\endequation
with the $E_{\eta_i}$ determined by the asymptotic expansion of $E_\lambda$ as
$$
	r_1E_{\eta_1}+r_2E_{\eta_2}~=~C
$$
\equation
	r_1E_{\eta_1}\eta_1+r_2E_{\eta_2}\eta_2~=~C\bar\phi+r_1+r_2-1
	\label{dp2}
\endequation
which allows the $E_{\eta_i}$ to be determined explicitly in terms of the mean
 value  $\bar\phi$ of $\phi$.
\equation
	\left(\matrix{{r_1E_{\eta_1}}\cr{r_2E_{\eta_2}}}\right)
	= \frac{1}{\eta_2-\eta_1} \left(\matrix{
	(\eta_2-\bar\phi)C+1-(r_1+r_2)\cr (\bar\phi-\eta_1)C+(r_1+r_2)-1}\right)
	\label{dp4}
\endequation
The solution to Eq. (\ref{dp1}) is simply
\equation
	E_\lambda=\frac{1}{2}\left( W^\prime(\lambda) -
	\sqrt{\left(W^\prime(\lambda)\right)^2-4\sum_{i=1}^2
	\frac{r_iE_{\eta_i}}{\lambda-\eta_i}}\right)
	\label{dp5}
\endequation
where
\equation
	W^\prime(\lambda)=\sum_{i=1}^2\frac{r_i}{\lambda-\eta_i}+C
	\label{dp6}
\endequation
is given by Eq. (\ref{womm}). Notice that we have chosen to write
the solution to the quadratic with a ``$-$'' rather than with a ``$\pm$''
and to discuss the two solutions in terms of the
possible branches of the square root.

The correct choice of the branch of the square root in Eq. (\ref{dp5}) is
very subtle, even in the case of the ordinary Penner Model (see
\cite{twodgrav}).
First of all we must choose the branch of the square root so that $E_\lambda$
satisfies the asymptotic condition (\ref{easym}): $E_\lambda \sim 1/\lambda$
as $\lambda\rightarrow\infty$. This will be satisfied provided the square root
is chosen to have no branch cuts going out to infinity and that
the square root approaches
$+C$ at infinity. There are however two other conditions which must also
be satisfied.
Notice from Eq. (\ref{dp4}) than $E_{\eta_i}$ is some finite number
so that $E_\lambda$ is not singular at the singularities of the potential.
We shall see that this
condition is not automatically satisfied by the solution (\ref{dp5}).
Furthermore we would like the ``Master Field'' $\Phi$
which is encoded by $E_\lambda$ to be a Hermetian
Matrix with $real$ eigenvalues. This requires the branch cuts to
be on the real axis with a purely imaginary discontinuity so that
the density of eigenvalues is real and positive. We shall see below that
it is often impossible to satisfy all these conditions simultaneously though
it may be possible to relax these conditions somewhat and still maintain
an interesting solution. In fact we shall see that despite the fact that
in the ordinary Double Penner model there are both one--cut and two--cut
solutions in the KM case in which the conditions (\ref{wommc}) must
be satisfied, only one--cut solutions will be possible (i.e. the second
cut will correspond to two degenerate branch points leading to
a zero density of eigenvalues).

Before proceeding to analyze the
branch cuts it is useful to write Eq. (\ref{dp5}) in an
alternate form. Note that the location of the
branch points are found by finding the zeros of the function under the
square root in Eq. (\ref{dp5}). This is a quartic equation for $\lambda$.
Let $\xi_1,\xi_2,\xi_3,\xi_4$ be the solutions to this quartic equation.
The requirement of a real, positive definite, density of eigenvalues will
require these roots to be real though there may be
some degeneracy among them. We thus expect, in general,
both one--cut and two--cut solutions to the Double Penner Matrix Model.
Let us assume, without loss of generality,
that $\xi_1\le\xi_2\le\xi_3\le\xi_4$
Eq. (\ref{dp5}) for $E_\lambda$ can now be written
\equation
	E_\lambda=\frac{1}{2}\left( W^\prime(\lambda) -
	\frac{C}{(\lambda-\eta_1)(\lambda-\eta_2)}
	\sqrt{(\lambda-\xi_1)(\lambda-\xi_2)(\lambda-\xi_3)
	(\lambda-\xi_4)}\right)
	\label{dp7}
\endequation
As discussed above, the asymptotic condition on $E_\lambda$ requires
the branch cuts of the square root to be chosen so that the square root
approaches $\lambda^2$ at infinity and that there are no branch cuts which
go out to infinity.

Implementation of the further conditions described above will be
discussed in Sec. \ref{ss3c}. Before doing so we should point out
that the density of eigenvalues which is extracted
from Eq. (\ref{dp7}) will always be normalized and have the correct value of
$\bar\phi$ provided only that $E_\lambda$ has no singularities and that
there are no cuts going out to infinity. To see this recall that $E_\lambda
\sim 1/\lambda$ as $\lambda\rightarrow\infty$. Thus
\equation
	\frac{1}{2\pi i}\oint_C E_\lambda ~ d\lambda~=~1
\endequation
where $C$ is a circle at infinity. If $E_\lambda$ has no additional
singularities then this integral can be written as an integral
over the discontinuities across the cuts of $E_\lambda$ which is just
the total normalization $\int \rho(\phi)d\phi$ which is thus equal to 1.
The mean value of $\phi$ can similarly be calculated. From the asymptotic
expansion (\ref{easym}) of $E_\lambda$ we see that
\equation
	\frac{1}{2\pi i}\oint_C \lambda E_\lambda ~ d\lambda~=~\bar\phi
	\label{dp9}
\endequation
which is now valid for the $E_\lambda$ in Eq. (\ref{dp7}). Recall however that
the
parameters $\xi_i$ in (\ref{dp7}) $depended$ on $\bar\phi$. The result
(\ref{dp9})
is valid, however, independently of this value of $\bar\phi$. Again, if
$E_\lambda$
has no additional singularities the contour can be deformed to a contour
surrounding the cuts so that $\int \phi\rho(\phi)d\phi=\bar\phi$ is this same
value of $\bar\phi$ which, as we recall, was extracted from the same
asymptotic expansion of $E_\lambda$.
(If the cuts are chosen so that $E_\lambda$ $is$ singular
at either one or both of the $\eta_i$ then both the normalization and
the mean value of $\phi$ will get an extra contribution from the singularity.
It is not difficult to show, for example, that the normalization will get a
contribution $r_i$ from a singularity at $\eta_i$.

\subsection{Review of Single Penner Case \label{ss3b}}

Before discussing the various possibilities in the Double Penner Case
let us review briefly how the cuts work in the ordinary Single Penner Model.
In this case the derivative of the potential is of the form
\equation
	W_p^\prime(\phi)=\frac{r}{\phi-\eta}+Q
\endequation
and the quadratic equation for $E_\lambda$ resulting from Eq. (\ref{onemme})
has the solution:
\equation
	E_\lambda=\frac{1}{2}\left\{\left(\frac{r}{\lambda-\eta}+Q\right)
	-\sqrt{\left(\frac{r}{\lambda-\eta}+Q\right)^2-4\frac{Q}{\lambda-\eta}}
	\right\}
	\label{sp2}
\endequation
with
\equation
	rE_\eta=Q
\endequation
The location $\xi_{\pm}$  of the branch points occurs when the square root
vanishes i.e.
\equation
	\xi_{\pm}=\eta+\frac{2-r}{Q}\pm \frac{2}{Q}\sqrt{1-r}
	\label{sp4}
\endequation
When $r>1$ there are no real solutions. This is the case when the potential
$W_p(\phi)$ has an attractive logarithmic singularity at $\eta$ with a strength
greater than 1. When $r<0$ the potential has a repulsive singularity which
results in a local minimum of $W_p$ which occurs at $\phi>\eta$ if $Q>0$
and at $\phi<\eta$ if $Q<0$. In this case it is easy to check that both
$\xi_{\pm}$ are on the same side of the singularity as this local minimum.
In the case $0<r<1$ the potential has an attractive singularity and $W_p$
has only a local maximum. In this case both branch points $\xi_{\pm}$ lie
side of the singularity opposite to this local maximum.

In this Single Penner case we can, as discussed above, try to implement
the three conditions: $E_\lambda\sim 1/\lambda$ as $\lambda\rightarrow\infty$;
$E_\lambda$ nonsingular at $\eta$; and $\rho(\phi)$ real and positive with
support on the real axis. The third condition implies that the branch points
and branch cuts must lie on the real axis. This guarantees that $\rho$
is real but not necessarily that it is positive. To analyze the other
conditions
we write, as in Eq. (\ref{dp7})
\equation
	E_\lambda=\frac{1}{2}\left\{\frac{r}{\lambda-\eta}+Q
	-Q\frac{\sqrt{(\lambda-\xi_+)(\lambda-\xi_-)}}{\lambda-\eta}\right\}
	\label{sp5}
\endequation
The asymptotic condition on $E_\lambda$ requires that there be no branch
cuts at infinity and that the square root be positive for $\lambda>\xi_\pm$.
This, together with the positivity requirement of $\rho$ implies that the
branch cut joins $\xi_\pm$ in a straight line along the real axis.
Furthermore the positivity of $\rho$ requires that $\xi_\pm>\eta$ if $Q>0$
and $\xi_\pm<\eta$ if $Q<0$. This is precisely the same result as
Eq. (\ref{sp4}) so that in the single penner case we are guaranteed that
the eigenvalue density will be positive if the branch cut joins the
$\xi_\pm$ in a straight line along the real axis.
{}From Eq. (\ref{sp2}) we can determine the behavior of the square root in
Eq. (\ref{sp5}) at the singularity up to a possible sign.
\equation
	E_\lambda\rightarrow \frac{1}{2}\frac{1}{\lambda-\eta}
	\left\{ r\mp\frac{Q}{\vert Q\vert}\frac{\vert r\vert}{r}r\right\}
	~~~~~~{\rm as}~\lambda\rightarrow\eta
\endequation
where the minus sign is used if the square root in (\ref{sp5}) is positive
which occurs if $\eta>\xi_\pm$
and the plus sign is used if it is negative which occurs if $\eta<\xi_\pm$.
Thus we establish the following condition for $E_\eta$ to be nonsingular:
If $\eta>\xi_\pm$ then the sign of $Q$ and $r$ must be the same
whereas if $\eta<\xi_\pm$ the $Q$ and $r$ must have opposite sign.
Thus, for the cancellation of singularities, we must have
\equation
	(\eta-\xi_\pm)Qr~>~0
\endequation

In the Single Penner Model the above condition is satisfied if and only
if $r<0$. This is evident from Eq. (\ref{sp4}) which implies that
$\xi_\pm>\eta$ if $Q>0$
whereas $\xi_\pm<\eta$ if $Q<0$. Thus both cases require $r<0$ for the
regularity of $E_\lambda$ at the singularity. This is physically reasonable
since this is precisely the case in which the potential has a local minimum.

The conventional way to analyze the case $1>r>0$ has been to relax the
positivity condition on the density of eigenvalues (see \cite{twodgrav} for
example)
and to allow the branch cut to go around the singularity so that the branch
cuts
goes from $\xi_-$ to $\xi_+$ by first going around $\eta$.  In the case
$1>r>0$ this will lead to a distribution of eigenvalues with complex support
but $E_\lambda$ will have both the correct asymptotics and it will be
nonsingular at $\eta$. One of the reasons this choice of cut is interesting
is that it behaves {\bf nearly} the same as a distribution of eigenvalues
with a delta function singularity (of strength $r$) at $\eta$  and
and the remainder of the eigenvalues along the real axis between the $\xi_\pm$.
This can be seen by noting that if we compute the average of any {\bf analytic}
function $f(\lambda)$
\equation
	I~=~\int_{\rm cut} d\lambda~ \rho(\lambda)f(\lambda)
\endequation
then this integral will be independent of the precise path which the
cut takes around the singularity. In fact by taking the cut from $\xi_-$
just below the real axis then around the singularity and back just above
the real axis to $\xi_+$ one can easily check that
\equation
	I~=~rf(\eta)+\int_{\xi_-}^{\xi_+}d\lambda~\hat\rho(\lambda)f(\lambda)
\endequation
with $\hat\rho$ normalized to $1-r$. This {\it looks as if} we could write
\equation
	\rho(\lambda)~=~ r\delta(\lambda-\eta)+\hat\rho(\lambda)\
\endequation
with $\hat\rho$ having support on $(\xi_-,\xi_+)$. This is often called
``condensation of eigenvalues'' since some fraction of the eigenvalues
``condense'' at the singularity. Unfortunately this interpretation is not
quite correct since if we were to evaluate the average of a {\it nonanalytic}
function $g(\lambda)$ (for example if we were to evaluate the Free Energy
which contains a logarithmic cut precisely in the region of interest) then
the result would be {\it dependent} on the precise path of the cut.
In fact, for a general path, the Free Energy is not even real though some
authors have chosen the contour in such a way that the Free Energy is real.

\subsection{Structure of Cuts in the Double Penner Case \label{ss3c}}

In the previous subsection we saw that even in the ordinary Penner model
proper one--cut solutions which satisfy all our
requirements do not necessarily  exist and there are cases when either
no solutions exist or when only very unusual solutions with
``eigenvalue condensation'' are present.
We now continue with the discussion in Subsection (\ref{ss3a}) of
the Double Penner Case where we shall find a similar situation.
Let us focus attention on Eqs. (\ref{dp5}) and (\ref{dp7}).
As in the Single Penner case the requirement
that $E_\lambda$ have no cut singularities at infinity and that $\rho(\lambda)$
be real and positive requires all the branch cuts to be on the real line.
Recalling  that
$\xi_1<\xi_2<\xi_3<\xi_4$ we must then have one cut from $\xi_1$ to $\xi_2$
and the other from $\xi_3$ to $\xi_4$.
It is also clear from Eqs. (\ref{dp5}) and (\ref{dp7})
that if $E_\lambda$ is to be regular
at the singularities they cannot be on a branch cut. Thus each singularity
is either below $\xi_1$, between $\xi_2$ and $\xi_3$ or above $\xi_4$.
Finally recall that the asymptotic condition on $E_\lambda$ requires
that the square root in Eq. (\ref{dp7}) must be positive for $\lambda>\xi_4$.
This in turn implies that it is negative between the branch cuts
and positive for $\lambda < \xi_1$.

We are now ready to study the possible singularity of $E_\lambda$ near the
$\eta_i$.
Near the singularity at $\eta_1$
\equation
	E_\lambda\rightarrow \frac{1}{2}\frac{1}{\lambda-\eta_1}
	\left\{ r_1\mp\frac{C}{\vert C\vert}\frac{\vert r_1\vert}{r_1}
	\frac{\vert \eta_1-\eta_2 \vert}{\eta_1-\eta_2}r_1\right\}
	~~~~~~{\rm as}~\lambda\rightarrow\eta_1
	\label{cdp1}
\endequation
(Near the singularity at $\eta_2$ simply replace $1\leftrightarrow 2$
everywhere
in Eq. (\ref{cdp1}).) The correct sign in the above equation depends on the
sign of the square root near the singularity. The
minus sign is to be used when the square root is positive
i.e if $\eta_1>\xi_4$ or $\eta_1<\xi_1$ (in this case we shall say that the
singularity is ``outside'' the cuts) and the plus sign must be used when
it is negative i.e. if $\xi_2<\eta_1<\xi_3$ (in which case we shall say that
the singularity is ``inside'' (i.e. between) the cuts).
If the minus sign is used the singularity cancels provided
$(\eta_1-\eta_2)Cr_1>0$ whereas if the plus sign is used the cancellation
occurs
if and only if $(\eta_1-\eta_2)Cr_1<0$.
Let us assume without loss of generality that
$\eta_2>\eta_1$.  (It is obvious that this can be done for the general
double Penner case but for the KM Model the requirement that $C>0$
makes this not so obvious. There are however symmetries which
relate the case $\eta>B$ to the case $\eta<B$ so that the assumption
$\eta_2>\eta_1$ is completely general.)

Our conclusion is then that:
\begin{eqnarray}
	Cr_1>0 &~~~~~~& \eta_1~{\rm inside} ~~ (\xi_2<\eta_1<\xi_3)~~~~
	(\sqrt{~~~}~~-)\cr
	Cr_1<0 &~~~~~~& \eta_1~{\rm outside} ~~(\eta_1<\xi_1~{\rm
or}~\xi_4<\eta_1)~~~~(\sqrt{~~~}~~+) \cr
	Cr_2>0 &~~~~~~& \eta_2~{\rm outside} ~~ (\eta_2<\xi_1~{\rm
or}~\xi_4<\eta_2)~~~~(\sqrt{~~~}~~+)  \cr
	Cr_2<0 &~~~~~~& \eta_2 ~{\rm inside} ~~ (\xi_2<\eta_2<\xi_3)~~~~
	(\sqrt{~~~}~~-)\cr
	\label{cdp2}
\end{eqnarray}
(If $\eta_1>\eta_2$ this would of course be reversed.) We shall consider
the case $C>0$ throughout.

Although the above conditions guarantee that $E_\lambda$ is nonsingular
it does not guarantee that $\rho$ is positive. The best way to see this is
to look at Eq. (\ref{dp7}). It is clear that if there are two cuts with
no singularity separating them or with both singularities separating them,
then one will have a positive $\rho$ and the other a negative $\rho$.
If we define the regions
I, II and III as the regions $\lambda<\eta_1$, $\eta_1<\lambda<\eta_2$ and
$\lambda>\eta_2$ respectively then  one cut must be in region II and
the other must be in region III.  There is however a loophole. If one
of the cuts is degenerate (i.e. $\xi_1=\xi_2$ or $\xi_3=\xi_4$)
then $\rho$ is zero along the degenerate cut and it does  not
matter if it is in the wrong region. We can thus have single cut solutions
as follows: Either the cut ({$\xi_1,\xi_2$}) is in region II  or the cut
 ({$\xi_3,\xi_4$}) is in either region I or III
with the other cut being degenerate.

We now proceed to discuss the various cases individually.

\subsection{The Case $r_1<0$ and $r_2<0$ \label{nn}}

The simplest situation occurs when both $r_1<0$ and $r_2<0$. This corresponds
to  a potential $W(\phi)$ which has two repulsive logarithmic singularities
and two local minima. One minimum is between the two singularities and the
other is above $\eta_2$ (since we have assumed that $C>0$).(We call this
the ``two up'' or ``2u'' potential.) In this case the condition (\ref{cdp2})
for a
nonsingular $E_\lambda$  is  that the branch cuts lie in the
same region as the minima of the potential namely in regions II and II.
 This is what we expect since
the eigenvalues of a Matrix Model
are expected to be distributed about the minima of the potential
as discussed in Sec. \ref{interpret}.
The only remaining question is whether the branch cuts which are a
solution to the quartic equation resulting from Eq. (\ref{dp5}) do
in fact lie in the correct place. It is possible to show that they always
do which also guarantees us that the density of eigenvalues is
everywhere positive. (Recall from the previous section that when
positivity of a double cut solution requires the cuts
to be in regions II and III when $C>0$.)

The solution to the quartic equation leading to the branch points
$\xi_1,\xi_2,\xi_3,\xi_4$ of $E_\lambda$ in Eq. (\ref{dp5}) is in general
quite complicated. The simplest way to see the types of solutions which
are possible is by a graphical method. Our goal is to solve the equation
\equation
	\left(\frac{r_1}{\lambda-\eta_1}+\frac{r_2}{\lambda-\eta_2}+C\right)^2
	-4\left(\frac{r_1E_1}{\lambda-\eta_1}+\frac{C-r_1E_1}{\lambda-\eta_2}
	\right)=0
	\label{cuu}
\endequation
with $r_1E_1$ and $r_2E_2$ given in Eq. (\ref{dp5}) in terms of $\bar\phi$.
Without loss of generality we may choose $\eta_1=0$ and call $\eta_2=\eta$.
Let us also call $r_1E_1=\delta$ so that $r_2E_2=C-\delta$. Eq. (\ref{cuu})
is thus equivalent to:
\equation
	\left[ r_1\left(\lambda-\eta_2\right)+r_2\left(\lambda-\eta_1\right)
	+C\left(\lambda-\eta_1\right)\left(\lambda-\eta_2\right)\right]^2
	=4(\left(C\lambda-\delta\eta\right)
	\left((\lambda-\eta_1\right)\left(\lambda-\eta_2\right)
	\label{cuv}
\endequation
The left hand side (LHS) has zeros precisely at the extrema of the
potential $W(\lambda)$ which we call $\lambda_1$ and $\lambda_2$.
In our present case ($r_1<0$, $r_2<0$) we have $0<\lambda_1<\eta<\lambda_2$.
We now sketch  both the LHS and the RHS of Eq. (\ref{cuv}) for various
values of $\delta$ (which, we recall, is related linearly to $\bar\phi$).
The LHS is a quartic with the two degenerate roots $\lambda_1$ and
$\lambda_2$ whereas the RHS is a cubic which goes like
$4C\lambda^3$ as $\lambda\rightarrow +\infty$ and with roots at
$\eta_1$, $\eta_2$ and $\delta\eta/C$. We now notice the following:

For $\delta<0$ there is always a pair of roots in region III ($\lambda
>\eta$) but there are {\bf never} any roots in region II.
Depending on the values of the various parameters there $may$
be a pair of roots in region I. These would however lead to a negative
density of eigenvalues and, correspondingly, the $\rho(\lambda)$ for
$\lambda>0$ would have $\int_{\lambda>0}\rho(\lambda)d\lambda>1$.
In case there are no roots in region I the two additional roots
are complex and, consequently, the density of eigenvalues for $\lambda>0$
will $not$ be normalized to 1.

When $\delta$ is slightly positive, the roots in region III persist
but there are no roots in region I nor in region II. At some critical
$\delta_{c1}$ which lies in the interval ($0,\lambda_1$) a pair of
roots begins to appear in region II. This pair of roots persists for all
$\delta>\delta){c1}$. At some critical $\delta_{c2}>\lambda_2$
the pair of roots in region III disappears.

Thus for all values of $\bar\phi$ for which $\delta_{c1}<\delta<\delta_{c2}$,
$E_\lambda$ has two branch cuts in regions II and III for which the
density of eigenvalues is positive, normalized to 1 and for which
$\int\phi\rho(\phi)d\phi=\bar\phi$ as required. For the two special
cases $\delta=\delta_{c1}$ and $\delta=\delta_{c2}$ two of the branch
points become degenerate and the above two--cut solution reduces to
a one--cut solution. Notice also that for all the above cases
$\rho(\lambda)\sim\vert\lambda-\xi\vert^{1\over 2}$ near any branch point
$\xi$.

The ``physical'' reason for the existence of this large class of classical
solutions can be seen by referring to the mechanical analogue problem discussed
in Sec. \ref{interpret}. Since there is an infinite barrier separating the
regions II and III we expect that a solution will exist with any number
$n_1$ of particles in region II and $N-n_1$ particles in region III.
The degenerate cases $\delta=\delta_{c1}$ and $\delta=\delta_{c2}$
correspond to the cases when all $N$ particles are either in
region I or in region II.

\subsection{The Case $r_1>0$ and $r_2>0$ \label{pp}}

The next case we consider is the case in which both $r_1$ and $r_2$
are positive. (This case could correspond to a KM Model provided the
conditions (\ref{wommc}) were satisfied.)

The next case we consider is the case in which both $r_1$ and $r_2$
are positive. (This case could correspond to a KM Model provided the
conditions in Eq. (\ref{wommc}) were satisfied.) In this case
the potential has no minimum but it has two maxima at points
$\lambda_1$ and $\lambda_2$ in regions I and II respectively.
We expect no real normalizable solutions in this case though there may be
cases in which ``eigenvalue condensation'' occurs as was discussed
in Sec. \ref{ss3b} for the Single Penner case.

Using the conditions given in Eq. (\ref{cdp2}) (and recalling that
$C>0$) we see that in order to avoid singularities of $E_\lambda$ at
$\eta_1$ and $\eta_2$ we must have one cut in region I and the other in region
II (i.e. in the same regions as the extrema of $W(\lambda)$). This,
unfortunately, leads to a negative density of eigenvalues in $both$
regions I and II. We thus conclude that there are no normalizable
one--cut or two--cut solutions in this case.

In order to examine the possibility of ``eigenvalue condensation''
we use the graphical approach discussed in Sec. \ref{nn}.
As in this previous section we assume without loss of generality
that $\eta_1=0$ and we call $\eta_2=\eta$ and we conclude as follows:
For many values of the parameters there will $not$ be four real
branch points. In case four real branch points exist there are
two possibilities. Either the two branch cuts are in regions
II and III or in regions I and III. Let us begin by discussing the
case when the cuts are in regions II and III. In this case $\rho(\lambda)$
is positive along both cuts but $E_\lambda$ is singular at $both$
$\eta_1$ and $\eta_2$ unless we deform both branch cuts in regions II
and III to surround the singularities $\eta_1$ and $\eta_2$ respectively.
Recall from the single Penner case that for the purposes of computing
averages of analytic functions we can choose a contour which
circles around each singularity but otherwise goes along the
real axis between ($\xi_1$ and $\xi_2$) and between ($\xi_3$ and $\xi_4$).
The singularities at $\eta_1$ and $\eta_2$ contribute an amount
$r_1$ and $r_2$ respectively to the normalization. We thus expect that
such solutions should exist only if $r_1+r_2<1$.

Finally we examine the possible case in which the cuts lie in regions
I and III. In this case $\rho(\lambda)$ is positive in region III but
negative in region I. Thus, in order to have a sensible solution, the
two branch points in region I must be degenerate ($\xi_1=\xi_2$) so that
the negative $\rho$ in region I is of no concern. We will however have
a singularity of $E_\lambda$ at $\eta_2$ unless the cut in region III
surrounds the singularity $\eta_2$. In this case the singularity will
contribute $r_2$ to the normalization so we expect such solutions
only if $r_2<1$.

\subsection{The Case $r_1>0$ and $r_2<0$ \label{pn}}

In the case $r_1>0$ $r_2<0$   the potential has a minimum in region III
and a maximum in region I. The condition (\ref{cdp2}) for the existence
of a nonsingular solution is that the cuts lie in regions I and III.
In this case the density of eigenvalues $\rho(\lambda)$ will be
positive in region III but negative in region I. It thus
follows that the branch points in region I must coincide ($\xi_1=\xi_2$).
It is easy to see using the graphical method described in the previous
sections that, independent of the parameters of the potential (provided
$r_1>0$ $r_2<0$ ) such a solution to our quartic equation always exists.
This leads in every instance to a normalizable distribution of
eigenvalues in region III for which the eigenvalues are distributed
about the minimum of the potential.

We now consider the more general possibility in which we allow
``eigenvalue condensation'' by relaxing the condition (\ref{cdp2}).
Depending on the values of the parameters in the potential, there
may be no real set of branch points or else the branch cuts
will be in regions II and III. This leads to a positive density
of eigenvalues along both cuts but $E_\lambda$ will be singular
at $\eta_1$ unless the branch cut in region II surrounds $\eta_1$.
Since this singularity will contribute $r_1$ to the normalization,
we only expect this solution to occur if $r_1<1$.

\subsection{The Case $r_1<0$ and $r_2>0$ \label{np}}

Finally we turn to the case $r_1<0$ $r_2>0$. In this case there are
four possibilities:

\equation
 (a)	{\hskip 4cm} 4\vert r_1\vert C\eta > \left(r_1+r_2-C\eta\right)^2
	\label{nosln}
\endequation
In this case $W(\lambda)$ has $no$ extrema at all.

\noindent In the remaining cases $4\vert r_1\vert C\eta <
\left(r_1+r_2-C\eta\right)^2$.

\equation
 (b)	{\hskip 6cm}	r_1+r_2>C\eta
	\label{slnb}
\endequation
$W(\lambda)$ has first a maximum then a minimum in region I.

\equation
 (c)	{\hskip 4cm}	-C\eta<r_1+r_2<C\eta
	\label{slnc}
\endequation
$W(\lambda)$ has first a minimum then a maximum in region II.

\equation
 (d)	{\hskip 6cm}	r_1+r_2<-C\eta
	\label{slnd}
\endequation
$W(\lambda)$ has first a maximum then a minimum in region III.

In the case (a) we expect no nonsingular solution. Whereas in the cases
(b),(c) and (d) we guess that if the minimum is not too shallow we
should get a normalizable solution distributed about this minimum.
The condition (\ref{cdp2}) for nonsingular solutions in this case allows
the branch cuts to be either both in region I or both in region II
or both in region III.

We begin by analyzing case (a) in which $W(\lambda)$ has no extrema.
In this case, the graphical method implies that the
 four real solutions to the quartic equations, if they
exist, may be either in regions I and III or in regions II and III.
It can further be shown that they cannot be in regions I and III.
Thus they can only be in regions II and III. We thus conclude, as
expected, that in this case we do not have a nonsingular solution.
However in case there are two pairs of roots in regions II and III,
the resulting branch cuts lead to a real eigenvalue distribution
for which $E_\lambda$ will be singular and $\eta_2$ unless the cut
in region III is deformed to circle around $\eta_2$. This will lead
to ``eigenvalue condensation'' at $\eta_2$ which can only occur if
$r_2<1$.

Next we look at case (b) for which $W(\lambda)$ has first a maximum
then a minimum in region I. In this case the quartic equation
has four roots in region I (for suitable values of $\delta$)
provided only that $(r_1+r_2-C\eta)^2+4r_1C\eta$ is not too small
(i.e. the minimum of $W$ is not too shallow). The cut connecting
$\xi_1$ and $\xi_2$ must be degenerate since it would lead to a
negative $\rho(\lambda)$. We thus have a normalizable, nonsingular
one--cut solution which can be seen (using, for example, the graphical
method) to lie around the minimum of $W$. A precisely analogous
situation occurs for cases (c) and (d). We have a normalizable, nonsingular
distribution of eigenvalues near the minimum of the potential provided
the minimum is not too shallow.

\section{The Kazakov--Migdal Model} {\label{s4}}

We saw in the previous section that the ordinary ``Two--Pole'' Penner
Model has a rich variety of one--cut, two--cut and singular solutions.
In this section we apply these results to the KM Model.
According to the results of Sec. \ref{sskmm} the solution to the
KM Penner Model with the potential $V(\lambda)$ given in
Eq. (\ref{vprime}) is related to an ordinary (non KMM) ``Two--Pole''
Penner Model with $W(\lambda)$ given in Eq. (\ref{womm}) provided
(\ref{wommc})
\equation
	r_1+r_2=1;~~~~~~~~~\eta_2-\eta_1=C
	\label{s41}
\endequation
(Recall that $r_1=q$, $\eta_1=\xi$ and $\eta_2=B$.)
For simplicity we shall call $r_1=r$ which is, of course, also equal to $q$.
The results of Sec. \ref{s3} can now be applied directly to the KM Model.

The most obvious conclusion we can draw immediately is that the case
$r_1<0$ $r_2<0$ is $not$ applicable to the KM Model. This was the
only case in the previous section which admitted a nonsingular
two--cut solution. We thus conclude that in the KM Double--Penner
Model we have at best only one--cut solutions.

Consider first the case $r>1$. In this case $r=r_1>1>0$ and
$r_2=1-r<0$. This case was discussed in Sec. \ref{pn}. For all such
values of $r$ and for all $C=\eta=\eta_2-\eta_1$ we have a
normalized one--cut solution in region III. In fact the degenerate
branch points $\xi_1=\xi_2<0$ and $\eta_2<\xi_3<\lambda_2<\xi_4$
where $\lambda_2$ is the location of the minimum of $W(\lambda)$.
The alternate solution in which there is condensation of eigenvalues
at the pole $\eta_1$ $cannot$ occur in the KM case since $r=r_1>1$.

It is instructive to examine the KM Potential in this case and then
to compare our results with the physical expectation described in
Sec. \ref{interpret}. From Sec. \ref{interpret} we recall that we are
interested not simply in the potential $V$ which appears in the KM
action but in $V(\lambda)-D\lambda^2$ (see Eq. (\ref{interp1})).
Choosing, without loss of generality, $\xi=0$ and calling
$B=\eta$ we have
\equation
	V^\prime(\lambda)-2D\lambda= \frac{r}{\lambda}
	+\frac{(2D-1)(r-1)}{\lambda-\eta}+\eta-2D\lambda
	\label{s42}
\endequation
\equation
	V(\lambda)-D\lambda^2=r{\rm log}(\lambda)
	+(2D-1)(r-1){\rm log}(\lambda-\eta) +\eta\lambda-D\lambda^2
	\label{s43}
\endequation

Noting that
\equation
	\eta\lambda-D\lambda^2=-D\left(\lambda-\frac{\eta}{2D}\right)^2
	+{\rm constant}
	\label{s44}
\endequation
we see that, unlike $W(\lambda)$ which has a maximum in region III,
$V(\lambda)-D\lambda^2$ has a $maximum$ in this region.
This is an example of the situation discussed in Sec. \ref{interpret}
in which the eigenvalue distribution has its support at the maximum
of a potential. The eigenvalues can ``straddle'' the maximum  due
to the long range attraction of the eigenvalues which is a result
of the integral over the Gauge Fields.

The next case we consider is when $r<0$. In this case $r_1=r<0$
and $r_2=1-r>1$. This situation was discussed in Sec. \ref{np}.
First note that there can be no eigenvalue condensation in this
case since $r_2>1$. Furthermore Eq. (\ref{nosln}) implies that if
\equation
	\vert r\vert > \frac{1}{4}\left(\eta-\frac{1}{\eta}\right)^2
\endequation
there are no real normalizable solutions. In fact even if $\vert r\vert$
is somewhat less than this limit, the minimum of the potential $W(\lambda)$
is too shallow to admit normalized solutions. When $\vert r\vert$ is
sufficiently small we can look for solutions by first noting the
location of the extrema of $W(\lambda)$ using Eqs. (\ref{slnb},\ref{slnc},
\ref{slnd}). If $\eta<1$ the requirements of case (b) of
Sec. \ref{np} is satisfied so $W(\lambda)$ has a minimum in region I.
It can be shown that in this case there are never four real roots
in region I. (One way to see that this has to be the case is by
noting that if such a solution did exist then $\bar\phi$ would have
to be in region I. This would imply that $\delta=\eta-\bar\phi>\eta$.
The graphical method of the previous section then shows that there
can be no roots in region I.)

If $\eta>1$ we satisfy the requirements of case (c) resulting in
a density of eigenvalues whose support is in region II. There is
however a critical value $\eta_{cr}>1$ of $\eta$
with the property that for $\eta<\eta_cr$ there is never a normalized
distribution of eigenvalues (i.e. one never has two cuts in region II
one of which is degenerate). For $\eta>\eta_{cr}$ a proper solution
exists provided $\vert r\vert < \vert r_c(\eta)\vert< (\eta-1/\eta)^2/4$.
The curve $r_c(\eta)$ can be computed with the aid of Eq. (\ref{dp7}).
Using the fact that $E_\lambda$ is nonsingular at both $\lambda=0$ and
at $\lambda=\eta$ and using the fact that $E_\lambda\sim 1/\lambda$
as $\lambda\rightarrow\infty$ one derives the following equation:
\begin{eqnarray}
	&r^2=\xi_1\xi_2\xi_3\xi_4\cr
	&(1-r)^2=(\eta-\xi_1)(\eta-\xi_2)(\eta-\xi_3)(\eta-\xi_4)\cr
	&1=\eta(\frac{\xi_1+\xi+2+\xi_3+\xi_4}{2}-\eta)\cr
	\label{threes}
\end{eqnarray}
The critical line will occur when three of the $\xi_i$ are degenerate.
This leaves three equations with only two $\xi$'s unknown and
allows us to determine $r$ in terms of $\eta$ with the result that
\equation
	r_c(\eta)=\frac{2+3\eta^{3/2}-\eta^2}{4}
\endequation
When $r<r_c(\eta)$ the KM potential
$V(\lambda)-D\lambda^2$ has a minimum in the support of $\rho$.
Notice that there is no case when $r<0$ for which the minimum of $W$
is in region III.

Finally we consider the case $0<r<1$. In this case $1>r=r_1>0$ and
$1>1-r=r_2>0$ which is discussed in Sec. \ref{pp}. There are no
nonsingular solutions in this case. Unlike the previous examples
this is a case in which the ordinary Penner model can have
``eigenvalue condensation'' since both $r_1$ and $r_2$ are $<1$.
In fact we found cases in which ``condensation'' occurred just
at $\lambda=\eta$ and cases in which is occured at $both$
$\lambda=0$ and $\lambda=\eta$. Unfortunately the KM Potential
$V-D\lambda^2$ has a $maximum$ at $\lambda=\eta$ and thus
eigenvalue condensation cannot occur at this point.  The only possible
explanation is that the procedure which lead from the Ordinary Penner Model
to the KM Penner model does not work when the eigenvalue distribution
is singular. This is not surprising, especially in light of the problems
with interpreting these as real distributions. We thus conclude that
for  the case $0<r<1$ there are no solutions at all.

In summary we see that we $never$ have eigenvalue condensation in the
KM Penner Model and that nonsingular normalized solutions exist
for all $\eta$ when $r>1$ and for $\vert r\vert < \vert r_c(\eta)\vert<
(\eta-1/\eta)^2/4$ when $r<0$ and $\eta>1$. The phase diagram for
this model is shown in Figure 1. There are several  interesting
critical lines. Along the line $\eta=0$ the potential
$W^\prime(\lambda)=1/\lambda$.
This is an single--pole Penner potential with a critical value of
the coupling ($r=1$) but without a linear term.
The line $r=1$ ($\eta>0$) yields a potential
$W^\prime(\lambda)=1/\lambda+\eta$.
This is again a critical single--pole Penner potential but this time
with a linear term. The line $r=0$ is also a critical Penner potential
though centered at $\lambda=\eta$ and with a linear term.
Finally there is the critical line $r=r_c(\eta)$ given by Eq. (\ref{threes})
at which
the potential in region II become sufficiently deep to admit
normalized solutions.

The behavior of physical quantities such as the
susceptibility
\equation
	\chi=-\frac{d^2}{dr^2}F(r)
\endequation
where $F$ is the Free Energy  near the various critical points or lines
is a subject of much interest in Matrix Models. In the case of the
KM Penner Model this was studied in great detail by Makeenko \cite{makpen}
who computed the susceptibility as well as the various critical
exponents of the model.

\vfill
\eject

\section {Discussion}{\label{s5}}

In this paper we have studied in detail the large $N$ solutions to
a Kazakov--Migdal Model with two logarithmic singularities
\equation
	V(\phi)=r~\log{\phi}+(2D-1)(r-1)\log{\phi-\eta}+\eta
\endequation
and the
solutions to the related problem of an ordinary Matrix Model
with the Double Penner Potential
\equation
	W(\phi)=r_1\log{\phi}+r_2\log{\phi-\eta}+C
\endequation
The ordinary Double Penner Model has
 a rich phase structure which includes regions in parameter space
in which there are one--cut solutions and two--cut solutions,
regions in which there are no solutions and regions in which
the solutions are singular with ``eigenvalue condensation'' at
the poles. The KM Penner Model on the other hand has either one--cut
solutions or no solutions at all. Its phase diagram is shown in Fig. 1.

The method employed in this paper, which was first used for the
KM Model in Ref. \cite{dms}, emphasizes the power of the
Path Integral in solving this difficult mathematical problem which
otherwise would have involved finding the extrema of Actions involving
the logarithm of the Itzykson--Zuber determinant in Eq. (\ref{phii}).
The techniques of Ref. \cite{dms} reduce the problem to solving
an ordinary quartic equation.

\centerline{\bf ACKNOWLEDGEMENTS}

We wish to thank Richard Szabo and Gordon Semenoff for their advice
and help.
This work was supported in part by the Natural Science
and Research Council of Canada. Their support is greatfully acknowledged.
Nathan Weiss wishes to thank the Department of Particle Physics at the
Weizmann Institute, where he is presently on leave, for their support.

\centerline{\bf FIGURE CAPTIONS}

\bigskip

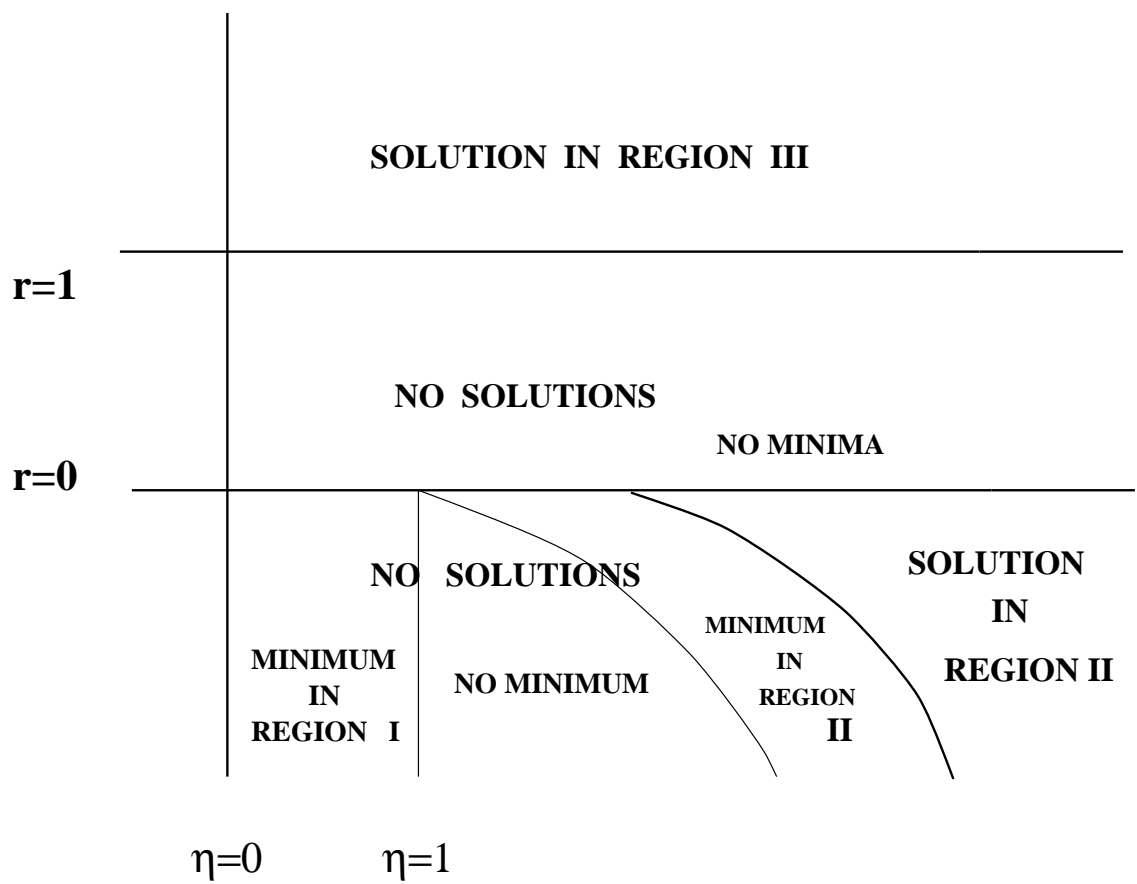
\noindent Figure 1: ~~ Phase Diagram for the Kazakov--Migdal Penner Model

\vspace{13cm}


\begin{references}

\bibitem{kmm}
V.A. Kazakov and A.A. Migdal, Nucl. Phys. {\bf B397}  214 (1993)
%
\bibitem{izform} Harish-Chandra, Amer. J. Math. {\bf 79} 87 (1957);
 	C.Itzykson and J.B. Zuber, J. Math. Phys. {\bf 21} 411 (1980);
 	M.L. Mehta, Comm. Math. Phys. {\bf 79} 327 (1981)
%
\bibitem{largen} E. Witten, {\it The $1/N$ Expansion in Atomic and
			Particle Physics}, in "Recent Developments in
			Gauge Theories " edited by t'Hooft et al;
			Plenum Press New York (1980)
%
\bibitem{gross}
D.J. Gross, Phys. Lett. {\bf B293} 181 (1992)
%
\bibitem{mig}
 A.A. Migdal, Mod. Phys. Lett.
	{\bf A8} 359 (1993); A.A. Migdal, Mod Phys. Lett. {\bf A8}
	153 (1993)
%
\bibitem{kmsw1}
I.I.Kogan, A.Morozov, G.W.Semenoff and N.Weiss,
Nucl. Phys. {\bf B395} 547 (1993)
%
\bibitem{GS92}
A.Gocksch and Y.Shen, Phys. Rev. Lett. {\bf 69}  2747 (1992)
%
\bibitem{KhM92}
S.B.Khokhlachev and Yu.M.Makeenko, Phys. Lett. {\bf B297} 345 (1992)
%
\bibitem{CAP92}
M.Caselle, A.D.'Adda and S.Panzeri,
Phys. Lett. {\bf B302} 80 (1993)
%
\bibitem{mig2}
A.A.Migdal, Mod. Phys. Lett. {\bf A8} 139 (1993)
%
\bibitem{M92}
Yu. M. Makeenko, Mod. Phys. Lett. {\bf A8}  209 (1993)
%
\bibitem{KM92}, S. Khokhlachev and Yu. Makeenko,
Mod. Phys. Lett. {\bf A7} 3653 (1992)
%
\bibitem{kmsw2}
I.I.~Kogan, A.Morozov, G.W.~Semenoff and N.~Weiss,
Int. J. Mod. Phys. {\bf A8} 1411 (1993)
%
\bibitem{mixed}
A.A.Migdal, Mod. Phys. Lett. {\bf A8} 245 (1993)
%
\bibitem{shat}
S.L. Shatashvili, Commun. Math. Phys. {\bf 154} 421 (1993)
%
\bibitem{moroz}
A. Yu. Morozov, Mod. Phys. Lett. {\bf A7} 3503 (1992)
%
\bibitem{dmsw} M. I. Dobroliubov,  A. Morzov, G.W. Semenoff, N. Weiss
	Int. J. Mod. Phys. {\bf A9} 5033 (1994)
%
\bibitem{kazak}V. A. Kazakov, Zh. Eksp. Teor. Fiz. 85, 1887 (1983)
[Sov. Phys. JETP] 58 1096 (1983).
%
\bibitem{ksw}
I.I.Kogan, G.W.Semenoff and N.Weiss, Phys. Rev. Lett. {\bf 69}
3435  (1992)
%
\bibitem{dms} M.I. Dobroliubov, Yu. Makeenko, G.W. Semenoff,
	Mod. Phys. Lett. {\bf A8} 2387 (1993)
%
\bibitem{makpen} Yu. Makeenko, Phys. Lett. {\bf B314} 197 (1993)\\
	Yu. Makeenko, ``Critical Scaling and Continuum Limits in the
	D$>1$ Kazakov--Migdal Model'', HEP-TH-9408029
%
\bibitem{mehta} Mehta, M. L. {\em Random Matrices} 2nd ed;
		Academic Press Inc. (1991)
%
\bibitem{twodgrav}
		S. Chauduri, H. Dykstra, H. , J. Lykken,
		Mod. Phys. Lett. {\bf A6} 1665 (1991) \\
		J. Ambjorn, C. F. Kristjansen, Yu. Makeenko,
		Phys. Rev. {\bf D50}  5193 (1994)
%
\bibitem{thesis}
 Lori Paniak, M.Sc. Thesis, University of British Columbia (1994)


\end{references}
\end{document}